\documentclass{eptcs}
\providecommand{\event}{GASCom 2026} 

\usepackage{iftex}
\usepackage{fetamont}
\usepackage[T1]{fontenc}

\usepackage{amsmath}
\usepackage{amssymb}
\usepackage{amsthm}
\usepackage{mathrsfs}
\usepackage{bm}
\usepackage{bbm}

\usepackage{comment}

\usepackage[svgnames,x11names]{xcolor}
\usepackage{hyperref}

\hypersetup{
colorlinks, 
linkcolor={SteelBlue},
citecolor={SteelBlue}, 
urlcolor={SteelBlue}
}

\usepackage{graphicx}
\usepackage{subfig}
\usepackage{tikz}
\usetikzlibrary{decorations.pathreplacing, tikzmark, calligraphy} 
\usepackage{tikz-cd}

\usetikzlibrary{tikzmark, arrows.meta, calc, bending}

\newtheoremstyle{slanted}
{}
{}
{\slshape}
{}
{\bfseries}
{.}
{ }
{}

\theoremstyle{slanted}
\newtheorem{theorem}{Theorem}[section]
\newtheorem{lemma}[theorem]{Lemma}
\newtheorem{corollary}[theorem]{Corollary}
\newtheorem{proposition}[theorem]{Proposition}
\newtheorem{definition}[theorem]{Definition}

\newtheorem{problem}{Problem}

\DeclareMathAlphabet{\mftm}{T1}{ffm}{m}{n} 
\newcommand{\A}{\mftm{A}} 
\newcommand{\B}{\mftm{B}}
\newcommand{\C}{\mftm{C}}
\newcommand{\D}{\mftm{D}}

\newcommand{\NN}{\mathbb{N}}

\DeclareMathOperator{\type}{type}
\DeclareMathOperator{\Types}{Types}
\DeclareMathOperator{\rec}{rec}

\def\ub#1{
\leavevmode\hbox{%
\setbox0\hbox{$#1$}\dp0 0pt
\vrule height.5ex width.4pt depth.33333ex \kern-.4pt
\vtop{\hbox{\kern.15em \box0\kern.15em}\kern.33333ex \hrule}%
\kern-.4pt \vrule height.5ex width.4pt depth.33333ex \kern-.4pt
}%
}

\def\ubb#1{
\leavevmode\hbox{%
\setbox0\hbox{$#1$}\dp0 0pt
{\color{white}\vrule height.5ex width.4pt depth.33333ex}\kern-.4pt
\vtop{\hbox{\kern.15em \box0\kern.15em}\kern.33333ex 
{\color{white}\hrule}}%
\kern-.4pt {\color{white}\vrule height.5ex width.4pt depth.33333ex}\kern-.4pt
}%
}

\definecolor{Green}{rgb}{0,0.6,0}
\definecolor{Orange}{rgb}{1.0, 0.5, 0.0}

\ifpdf
\usepackage[T1]{fontenc}        
\else
\usepackage{breakurl}           
\fi

\title{Frequency of Patterns in Smooth Sequences \\ 
over the Alphabet $\{1,3\}$}
\author{Damien Jamet
\institute{Universit\'e de Lorraine, CNRS, LORIA, \\UMR 7503, F-54000 Nancy, France}
\email{damien.jamet@loria.fr}
\and
Irène Marcovici
\institute{Univ Rouen Normandie, CNRS, Normandie Univ,\\ LMRS UMR 6085, F-76000 Rouen, France}
\email{irene.marcovici@univ-rouen.fr}
\and
Léo Poirier
\institute{Aix-Marseille Universit\'e, CNRS, \\
I2M UMR 7373, F-13000 Marseille, France}
\email{leo.poirier@univ-amu.fr}
\and
Thierry de la Rue
\institute{Univ Rouen Normandie, CNRS, Normandie Univ,\\ LMRS UMR 6085, F-76000 Rouen, France}
\email{thierry.de-la-rue@univ-rouen.fr}
}
\def\titlerunning{Frequency of Patterns in Smooth Sequences over the Alphabet $\{1,3\}$}
\def\authorrunning{D.~Jamet, I.~Marcovici, L.~Poirier \& T.~de~la~Rue}

\begin{document}
\maketitle

\begin{abstract}
We provide an ergodic theory framework to study statistical properties of smooth sequences over the odd alphabet $\{1, 3\}$. The arithmetic nature of this alphabet yields a partition of the subshift of smooth sequences based on their local structure, defining a notion of type for those sequences. We describe the substitutive structure of the smaller subshifts obtained by fixing the sequence of types of the successive derivatives of smooth sequences, from which we obtain the unique ergodicity of all these subshifts.  	    A direct consequence is that the asymptotic frequency of any finite pattern in a smooth sequence over $\{1,3\}$ is always well-defined and depends on its type sequence. Finally, we characterize the minimality of these subshifts. 
\end{abstract}

\section{Introduction}

The study of combinatorial structures arising from run-length encoding has been a subject of enduring interest in discrete mathematics. For many years, the canonical reference for this object has been the problem posed by Kolakoski in 1965~\cite{Kol65}, asking for a rule describing the self-reading sequence $\mathbb{K}=1221121221\dots$. However, this sequence was introduced and extensively investigated much earlier by Oldenburger in 1939~\cite{Old39}. The \textbf{Oldenburger--Kolakoski sequence} $\mathbb{K}$, defined over $\{1, 2\}$, is the unique fixed point of the run-length encoding operator $\Delta$ (which we call the \emph{derivative} operator) that starts with~$1$.

While its definition is elementary, the dynamical properties of $\mathbb{K}$ remain largely unknown. Dekking~\cite{Dek80} highlighted two fundamental open problems: \emph{mirror invariance} (a word is a factor iff its reversal is) and \emph{recurrence}. He established that mirror invariance implies recurrence and is equivalent to stating that every smooth word appears as a factor of $\mathbb{K}$. Regarding complexity, he conjectured a polynomial growth $p_\mathbb{K}(n) \asymp n^\rho$ with $\rho \approx 2.71$ \cite{Dek81}. Recent progress by Cassaigne and Henry \cite{cassaigne2026} provides new bounds for this complexity and formalizes the class of \emph{f-smooth} words. Other conjectures, such as the frequency of the symbol $1$ being $1/2$ \cite{Kea91}, also remain open, with current rigorous bounds at $[0.49992, 0.50008]$ \cite{Rao12}.

To gain insight into such self-descriptive structures, the notion of \emph{smooth sequences} (or $C^\infty$-sequences) was introduced as a natural generalization \cite{BL2003, BDLV06}. A sequence $x$ over a finite alphabet $\Sigma \subset \mathbb{N}$ is said to be \textbf{differentiable} if its derivative $\Delta x$ (the sequence of lengths of its consecutive mono-symbol blocks) remains a sequence over $\Sigma$. It is \textbf{smooth} if it is infinitely differentiable, i.e., $\Delta^k x \in \Sigma^{\mathbb{N}}$ for all $k \ge 0$. 

Research on smooth sequences is traditionally categorized by the parity of the alphabet.
On a two-letter alphabet, when symbols share an \emph{even} parity, the structure is perfectly balanced, and symbol frequencies are trivially $1/2$. Odd alphabets such as $\{1,3\}$ exhibit intermediate behavior: for instance, Baake and Sing \cite{BaakeSing2004} showed that the Kolakoski-(3,1) sequence relates to model sets with pure point diffraction spectra, providing a structural bridge between these sequences and quasicrystals.

In the present paper, we also focus on the dynamical and statistical properties of smooth sequences over $\Sigma = \{1,3\}$. While specific extremal cases have been studied \cite{BJP08}, the general distribution of frequencies in this setting remains poorly understood.

\paragraph{Main Results --}
To address these questions, we extend the study of smooth sequences to the bi-infinite setting. This transition is essential for ergodic approach, as it allows us to work within a compact, shift-invariant space: the subshift $X \subset \{1,3\}^\mathbb{Z}$ of bi-infinite smooth sequences. Our analysis relies on two main tools:
\begin{enumerate}
\item A \textbf{recoding} of smooth sequences over a four-letter alphabet $\mathcal{A} = \{\A, \B, \C, \D\}$ based on the four possible mono-symbol blocks.
\item A \textbf{classification} of these sequences into two distinct \emph{types} (0 and 1) based on their local structure, which is preserved under derivation.
\end{enumerate}
This framework reveals that $X$ decomposes into a disjoint union $X = \bigsqcup_{\tau \in \{0,1\}^\mathbb{N}} X_\tau$, where each $X_\tau$ is defined by a fixed sequence of types $\tau$. Our first main result is that each subshift $X_\tau$ (and its recoded counterpart $Y_\tau$) is uniquely ergodic (Corollary~\ref{cor::uniqueErgodicity}). As a direct consequence of this unique ergodicity, the asymptotic frequency of any finite pattern $w$ exists and is well-defined for both bi-infinite and one-sided smooth sequences (Theorem~\ref{thm::frequencies}). Furthermore, we provide a complete characterization of the minimality of these subshifts and we establish that $X_\tau$ (resp. $Y_\tau$) is minimal if and only if $\tau$ contains infinitely many zeros (Theorem~\ref{thm:minimal}). 

\section{Smooth sequences over $\{1, 3\}$ and their recodings}
\label{sec:def}
    
    \paragraph{Bi-infinite smooth sequences over ${\{1,3\}}$ --} The operator $\Delta$ maps any non-stationary bi-infinite sequence $x$ to the sequence of lengths of its consecutive mono-symbol blocks. To properly index this derived sequence, we adopt the following convention: the value $(\Delta x)_0$ at the origin of the derivative is the length of the mono-symbol block in $x$ that contains the term $x_0$. 
    
    \noindent Having defined the action of the derivative operator $\Delta$ on bi-infinite sequences, we now formalize the concepts of differentiability and smoothness in this setting:

    \begin{definition}
    	A bi-infinite sequence $x \in \{1,3\}^\mathbb{Z}$ is said to be \emph{differentiable} if its derivative $\Delta x$ is well-defined (i.e., $x$ is non-stationary) and belongs to $\{1,3\}^\mathbb{Z}$. Furthermore, $x$ is said to be \emph{smooth} if it is infinitely differentiable.
    \end{definition}
    \noindent Let us denote by $X \subset \{1,3\}^\mathbb{Z}$ the set of all bi-infinite smooth sequences. It is obviously invariant under the shift map $S$. Furthermore, $X$ is a closed subset of the product space $\{1,3\}^\mathbb{Z}$. As a closed and shift-invariant subset of a compact space, $X$ constitutes a compact subshift, providing the exact topological framework required for our ergodic study.
    
    \paragraph{Recodings of smooth sequences --} Since the derivative of a smooth sequence is also a sequence over the same alphabet, any $x \in X$ is necessarily a concatenation of blocks of lengths $1$ and $3$. 
    Consequently, a smooth sequence is composed of only four types of mono-symbol blocks, which we recode using the alphabet $\mathcal{A} = \{\A, \B, \C, \D\}$ defined as follows: ${\A = 1, \quad \B = 3, \quad \C = 111, \quad \D = 333}$.
    The \emph{recoding} of a smooth sequence $x \in X$ is the sequence $y = \texttt{rec}(x) \in \mathcal{A}^\mathbb{Z}$ where each term $y_i$ encodes a consecutive mono-symbol block (or run) in $x$. By convention, $y_0$ is the symbol encoding the block that contains $x_0$. We denote by $Y = \rec(X) \subset \mathcal{A}^\mathbb{Z}$ the subshift of all such recodings.
    An example of this correspondence is shown below, where the underlined symbol indicates the zero position of the sequence:
    \[\arraycolsep=2pt
        \begin{array}{lcrccccccccl}
        	x & = & \cdots 333 &  1\underline{1}1  &  333  &   1 &  3 &  1  &  333  &  111  &  333  &  1\cdots \\
            y = \text{rec}(x) & = & \cdots \D \; & \; \underline{\C} \; &  \; \D \; &   \A &  \B &   \A &  \; \D \; &  \; \C  \; &  \; \D \;&   \A \cdots
        \end{array}
    \]
    \noindent Observe that each symbol in the recoding sequence $y = \rec(x)$ corresponds to a single symbol in the derivative $\Delta(x)$: a $1$ in $\Delta(x)$ gives rise either to an $\A$ or to a $\B$ in $y$, whereas a $3$  in $\Delta(x)$ gives rise either to a $\C$ or to a $\D$ in $y$. Now, the structuring of $\Delta(x)$ into mono-symbol blocks yields a central property of $y = \rec(x)$: its decomposition into so-called \emph{elementary blocks}, described in the following array.
    \[
    \begin{array}{ccc}
    \text{mono-symbol block in }\Delta(x) & \text{corresponding factor in }x & \text{elementary block in }y \\
     1  & 1 \text{ or }3 &  \A \text{ or } \B \\
     3  & 111 \text{ or }333 &  \C  \text{ or } \D  \\
     111  & 131 \text{ or }313 &  \A\B\A  \text{ or } \B\A\B  \\
     333  & 111333111 \text{ or }333111333 &  \C\D\C  \text{ or } \D\C\D  
    \end{array}
    \]
    \noindent Note that these elementary blocks are either factors of $y$ of length 3, of the form $\A\B\A$, $\B\A\B$, $\C\D\C$, $\D\C\D$, or single letters $\A$, $\B$, $\C$, $\D$ that are not part of larger elementary blocks. Observe also that, since mono-symbol blocks in $\Delta(x)$ have length at most 3, elementary blocks in $y$ cannot overlap, and therefore there is a unique way of decomposing $y$ as a bi-infinite concatenation of elementary blocks.
    
    \noindent We can then naturally define a notion of derivation for recodings as $\Delta(rec(x)) = rec(\Delta x)$. In practice, using the elementary block decomposition of any $y \in Y$, this results in the following local rules:
    \begin{equation*}
    	\Delta : \left| \begin{array}{lcl}
    		\A\B\A, \B\A\B & \longmapsto & \C \\
    		\C\D\C, \D\C\D & \longmapsto & \D \\
    		\A, \B & \longmapsto & \A \\
    		\C, \D & \longmapsto & \B
    	\end{array} \right.
    \end{equation*}    

\paragraph{Two types of smooth sequences and recodings --} The above local rules give two possible preimages for  mono-symbol blocks in $\Delta(x)$. As it turns out, the arithmetic nature of the alphabet $\Sigma = \{1, 3\}$ imposes a strict parity constraint on the positions of the blocks, which implies that the choice of the preimage of one mono-symbol block determines all the others in the process of taking the primitive. This yields the following proposition. 

\begin{proposition}
	\label{prop:type}
	A recoding sequence $y \in Y$ cannot simultaneously contain an elementary block in $\{\A\B\A, \D\C\D\}$ and an elementary block in $\{\B\A\B, \C\D\C\}$.
\end{proposition} 

This internal consistency allows us to classify smooth sequences into two distinct categories based on their local structure.
\begin{definition}
  	Let $x \in X$ be a smooth sequence and $y = \rec(x)$ its recoding. We say that $x$ and $y$ are of \emph{type~0} if  $y$ contains an elementary block in $\{\A\B\A, \D\C\D\}$. Conversely, we say that $x$ and $y$ are of \emph{type~1} if $y$ contains an elementary block in $\{\B\A\B, \C\D\C\}$. 
  	
  	We denote the type of $x$ and $y$ by $\type(x) = \type(y) \in \{0, 1\}$. The sequence of types of all iterated derivatives is denoted by:
  	\[ \Types(x) = \Types(y) := \bigl(\type\left(\Delta^n x\right)\bigr)_{n \in \mathbb{N}} \in \{0, 1\}^\mathbb{N}. \]
\end{definition}
This classification gives us a way to partition $X$ and $Y$ by specifying the sequence of types of their elements.
\begin{definition}
      For a sequence of types $\tau \in \{0, 1\}^\mathbb{N}$, we define $X_\tau$ as the set of smooth bi-infinite sequences whose type sequence is exactly $\tau$:
  	\[ X_\tau := \{ x \in X \mid \Types(x) = \tau \}. \]
  	The set of corresponding recodings is denoted by ${Y_\tau := \rec(X_\tau) = \{ y \in Y \mid \Types(y) = \tau \}}$.
\end{definition}

$X_\tau$ and $Y_\tau$ are clearly closed and shift-invariant. We also prove that they are always nonempty: therefore, they are subshifts over respectively the alphabets $\{1, 3\}$ and $\{\A, \B, \C, \D\}$.

\paragraph{Substitutive structure --} We now define two substitutions that will, given the type of a recoding $y \in Y_\tau$ and its derivative $\Delta y$, help us reconstruct back $y$ from local rules:
\[
\varphi_0:\begin{cases}\A \mapsto \A \\ \B \mapsto \D \\ \C \mapsto \A\B\A \\ \D \mapsto \D\C\D \end{cases}     
\qquad      
\varphi_1:\begin{cases}\A \mapsto \B \\ \B \mapsto \C \\ \C \mapsto \B\A\B \\ \D \mapsto \C\D\C \end{cases}.     
\]
The action of $\varphi_0$ and $\varphi_1$ are extended by morphism to bi-infinite sequences: for $t = 0$ or $1$, applying $\varphi_t$  on every letter of $y \in Y$ yields the concatenated sequence of blocks 
\[ \dots \varphi_t(y_{-1}) \varphi_t(y_0) \varphi_t(y_1) \varphi_t(y_2)\dots \] 
To properly index the resulting sequence, we adopt the convention that the block $\varphi_t(y_0)$ begins exactly at index $0$. The positions of all other blocks are then uniquely determined by expanding outwards to the left and right.  These two substitutions act as right inverses of the derivative operator $\Delta$: for any $y \in Y$ and any $t \in \{0,1\}$, we have $\Delta(\varphi_t(y)) = y$. Conversely, applying the substitution  $\varphi_{\type(y)}$ to the derivative $\Delta y$ recovers the original sequence $y$ only up to a bounded shift.
In fact, for any $L \geq 1$, given the $L$ first terms of the sequence of types of a recoding $y$, we can recover it from its $L$-derivative $\Delta^L y$.
\begin{proposition}
\label{Prop prim}
For all $\tau \in \{0,1\}^\NN$, for all $y \in Y_\tau$, for all $L \ge 1$, there exists $i \in \{0,\dots,3^L-1\}$ such that 
\begin{equation}
    \label{eq:Lsubstitutions}
    y = S^i \varphi_{\tau_0} \circ \varphi_{\tau_1} \circ \cdots \circ \varphi_{\tau_{L-1}} \bigl( \Delta^L y \bigr).
\end{equation}
\end{proposition}

\section{Main results}\label{sec:MainResults}

\subsection{Frequency of patterns and unique ergodicity of $X_\tau$ and $Y_\tau$}

Let $\Sigma$ be a finite alphabet, and let $w \in \Sigma^*$ be a finite word. An \emph{occurrence} of $w$ in a sequence $x \in \Sigma^* \cup \Sigma^{\mathbb{N}} \cup \Sigma^{\mathbb{Z}}$ is an index $i$ such that $x_{i+k} = w_k$ for all $0 \le k < |w|$. We denote by $|x_0 \cdots x_{n-1}|_w$ the number of occurrences of $w$ in the prefix of length $n$ of $x$.
The \emph{frequency (of occurrence)} of $w$ in $x$ is defined as the limit:
\[
\mathrm{freq}_x(w) := \lim_{n \to \infty} \frac{|x_0 \cdots x_{n-1}|_w}{n},
\]
provided that this limit exists.

\begin{theorem}\label{thm::frequencies}
For any $\tau \in \{0, 1\}^\NN$, we have the following results:
\begin{itemize}
\item $\forall y \in Y_\tau$, $\forall w \in \{\A,\B,\C,\D\}^*$, the frequency of $w$ in $y$ exists, and depend only on $\tau$ and $w$ ;
\item $\forall x \in X_\tau$, $\forall w \in \{1, 3\}^*$, the frequency of $w$ in $x$ exists, and depend only on $\tau$ and $w$.
\end{itemize}
\end{theorem}

We recall that a subshift is said to be \emph{uniquely ergodic} if there is a unique shift-invariant probability measure supported on this subshift. A reformulation of the above result is the following:

\begin{corollary}\label{cor::uniqueErgodicity}
The subshifts $X_\tau$ and $Y_\tau$ are uniquely ergodic
\end{corollary}

We denote by $\mu_\tau$ and $\nu_\tau$ the respective unique shift-invariant probability measure on $X_\tau$ and $Y_\tau$.

\subsection{Sketch of the proof}
\label{sec:uniqueErgodicity-Yt}
We first establish the result for $Y_\tau$, by exploiting its substitutive structure. Then we show how $X_\tau$ inherits the unique ergodicity property from $Y_\tau$. 

We recall that, as for any subshift over a finite alphabet, the set of Borel probability measures over $Y_\tau$ is nonempty, and the topology of weak convergence of probability measures turns this set into a metrizable compact space. For $y\in Y_\tau$ and $n \ge 1$, we define the \emph{empirical measure}
\[
m_y^n := \frac{1}{n} \sum_{j = 1}^n \delta_{S^j y}. 
\]
By compacity, there exists an increasing sequence $(n_k)$ along which the sequence of empirical measure converges to an accumulation point $\nu$. This accumulation point is always shift-invariant. Moreover, this weak convergence amounts to saying that the frequency of any finite word $w \in \{\A, \B, \C, \D\}^*$ exists along $(n_k)$, and is given by $\nu([w])$, the measure of the cylinder set associated to the word $w$.
The core of the argument consists in proving that this frequency depends neither on the choice of the increasing sequence $(n_k)$ along which the convergence holds, nor on the choice of $y \in Y_\tau$. 

A preliminary step shows that, for all $L \ge 0$, there exists an increasing sequence of integers $(n_k^L)_k$  (depending on $(n_k)$ and $y$) along which we also have the weak convergence of the empirical measures $m_{\Delta^L y}^n$. We can then consider $a^L$, $b^L$, $c^L$ and $d^L$ to be the respective frequencies of the symbols $\A$, $\B$, $\C$ and $\D$ in $\Delta^L y$ measured along $(n_k^L)$. By alternance of the blocks of $1$s and the blocks of $3$s in any smooth sequence, we easily see that 
\[ a_L + c_L = b_L + d_L = 1/2. \] Therefore we can focus on the study of $a^L$ and $b^L$ (from which the other two can be deduced).
The substitutive structure of $Y_\tau$ yields the following result:
    
    \begin{lemma}
        \label{lemma:h0h1}
        For any $L \geq 1$, if $\tau_L = i \in \{0, 1\}$, then
        ${ \left(a^L,b^L\right) = h_i\left(a^{L+1},b^{L+1}\right) }$, where $h_0$ and $h_1$ are the following homographies:
    \[
    h_0(a, b) := \frac{1}{3 - 2(a + b)} \left(1 - a, \frac{1}{2} - a\right), \qquad h_1(a, b) := \frac{1}{3 - 2(a + b)} \left( \frac{1}{2} - a, 1 - a\right).
    \]
    \end{lemma}
    
    \begin{figure}[ht!]
     \begin{center}
            \includegraphics[width = 4cm]{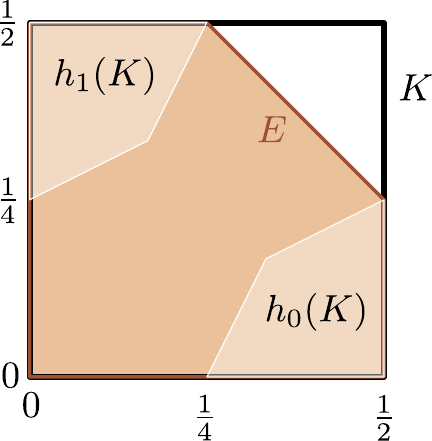}                                           
     \end{center}
    \caption{Homographies $h_0$ and $h_1$ acting on the square $K = [0,\frac{1}{2}] \times [0,\frac{1}{2}]$}\label{fig:h0h1}
    \end{figure} 
    
    As can be observed on Figure \ref{fig:h0h1}, taking Lemma \ref{lemma:h0h1} into account, their respective actions on the square $K$ imply that any $(a^L,b^L)$ is actually in ${E := \{(a,b) \in K \; | \; a+b \leq 3/4\}}$, so we consider $h_0$ and $h_1$ as maps: $E \to E$. Seen as such, both homographies can be shown to be contracting for $\|\cdot\|_\infty$, which implies the following lemma
    
    \begin{lemma}
     For every sequence $t = (t_L)_{L \ge 0} \in \{0, 1\}^\NN$, there exists $\bigl(a(t), b(t)\bigr) \in E$ such that 
     \[ \bigcap_{L \ge 0} h_{t_0} \circ \cdots \circ h_{t_L} (E) = \bigl\{\bigl(a(t), b(t)\bigr)\bigr\}. \]
    \end{lemma}
    
    Coming back to the frequencies of letters observed in $\Delta^L y$ along the increasing sequence $(n_k^L)_{k\ge1}$, we can conclude using Lemma~\ref{lemma:h0h1} that these frequencies are dependent neither on the choice of $(n_k)$ nor on that of $y$, they depend only on $\tau$:
    
    \begin{corollary}
    \label{cor:symbol_frequency}
     For each $L\ge 0$, we have 
     \[
      a^L = a({S^L \tau}), \quad
      b^L = b({S^L \tau}), \quad
      c^L = \frac{1}{2} - a({S^L \tau}), \quad \text{ and }\quad
      d^L = \frac{1}{2} - b({S^L \tau}).
     \]
    \end{corollary}

    Once we know that the frequencies of symbols in all the derivatives of $y$ depend only on $\tau$, we exploit again the substitutive structure of $Y_\tau$ to prove that similar results hold for the frequencies of larger patterns measured along $(n_k)$, and the unique ergodicity of $Y_\tau$ follows.
    
    \paragraph{Coming back to $X_\tau$ --}
    
    To come back to the subshift $X_\tau$ of smooth sequences of Types $\tau$, we establish a conjugacy between $Y_\tau$ and the system induced by $X_\tau$ on some subset $B\subset X_\tau$. By isomorphism, the system $(B, S_B)$ (where $S_B$ is the transformation induced on $B$ by the shift map $S$) is also uniquely ergodic. Then the unique ergodicity of $X_\tau$ follows.

\section{Minimality}

\label{sec:minimal}
Once we know that $(Y_\tau, S)$ and $(X_\tau, S)$ are uniquely ergodic for all $\tau$, the question of the minimality of these systems arises naturally. We recall the following classical result (see \emph{e.g.} \cite[Theorem~6.17]{Walters1982}): if $(X, T)$ is a uniquely ergodic topological dynamical system, with unique invariant probability measure $\mu$, then $(X, T)$ is minimal if and only if $\mu(U) > 0$ for all non-empty open set $U\subset X$. In the context of a symbolic dynamical system, this condition is equivalent to saying that the unique invariant measure charges every cylinder $[w]$ for all finite word $w$ in the language of the subshift. We have therefore to study whether, for a given $\tau\in\{0,1\}^\NN$, the following is true: 
\begin{equation}
\label{eq:minimality_condition}
\forall w\in \mathcal{L}(Y_\tau),\ \nu_\tau([w]) > 0.
\end{equation}

\begin{theorem}\label{thm:minimal} 
For all $\tau \in \{0,1\}^\mathbb{N}$, the subshift $(Y_\tau,S)$ is minimal if and only if $\tau$ has infinitely many $0$s
\end{theorem}  

\paragraph{Non-minimal cases --} By looking back at the definition of $\varphi_1$, the only way to produce a symbol $\D$ through this substitution is by applying it to some already-present $\D$: $\varphi_1(\D) = \C\D\C$. This results in the fact that whenever there exists an $L \geq 0$ such that $S^L \tau = 1^\infty$, the corresponding subshift $Y_\tau$ will contain words that have a finite amount of occurrences of the symbol $\D$ in them, contradicting \eqref{eq:minimality_condition} because in that case, $\D \in \mathcal{L}(Y_\tau)$ and $\nu_\tau([\D]) = 0$.

\paragraph{Other cases are minimal --} On the other hand, a closer study of the actions of $h_0$ and $h_1$ on the square $K$ yields a characterisation of the respective cases where each symbol frequency equals zero. Then, for a larger word $w$, we can always find a strictly shorter word $w'$ which occurrence can be associated with that of $w$; hence the ability to inherit minimality from the strict positivity of the frequencies of symbols.

\section{Perspectives}

\paragraph{Study of the set of possible frequencies --} 

Given a sequence type $\tau$, we recall the notation $a = a(\tau) := \nu_\tau([\A])$, and likewise for $\B,\C$ and $\D$. By the unique ergodicity of $(X_\tau,S)$, all the sequences of $X_{\tau}$ have the same frequency of $1$s. We denote by $f_{\tau} :=\mu_\tau([1])$ the value of this frequency.

\begin{figure}[!h]%
        \centering
        \subfloat[\centering  $1$ iteration]{{\includegraphics[width=.24\textwidth]{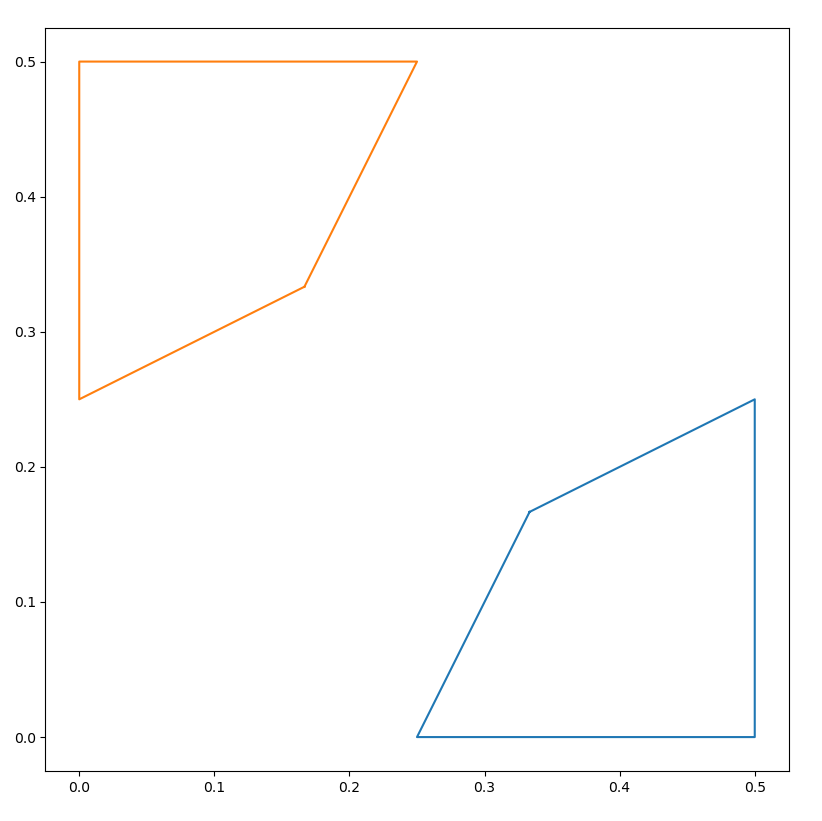}}}%
        \subfloat[$2$ iterations]{{\includegraphics[width=.24\textwidth]{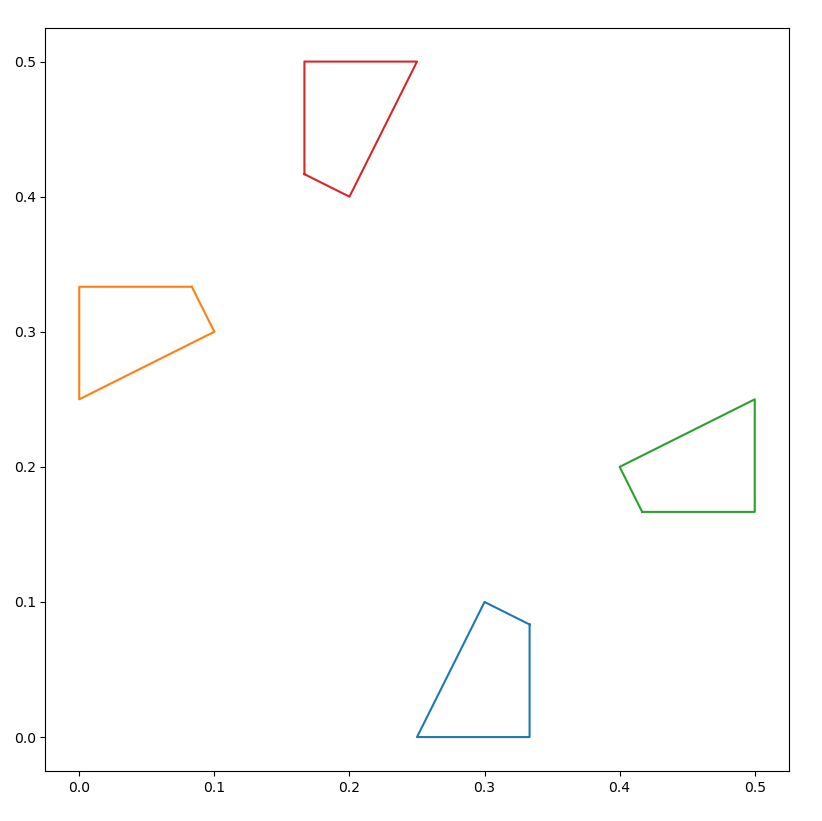}}}
        \subfloat[$4$ iterations]{{\includegraphics[width=.24\textwidth]{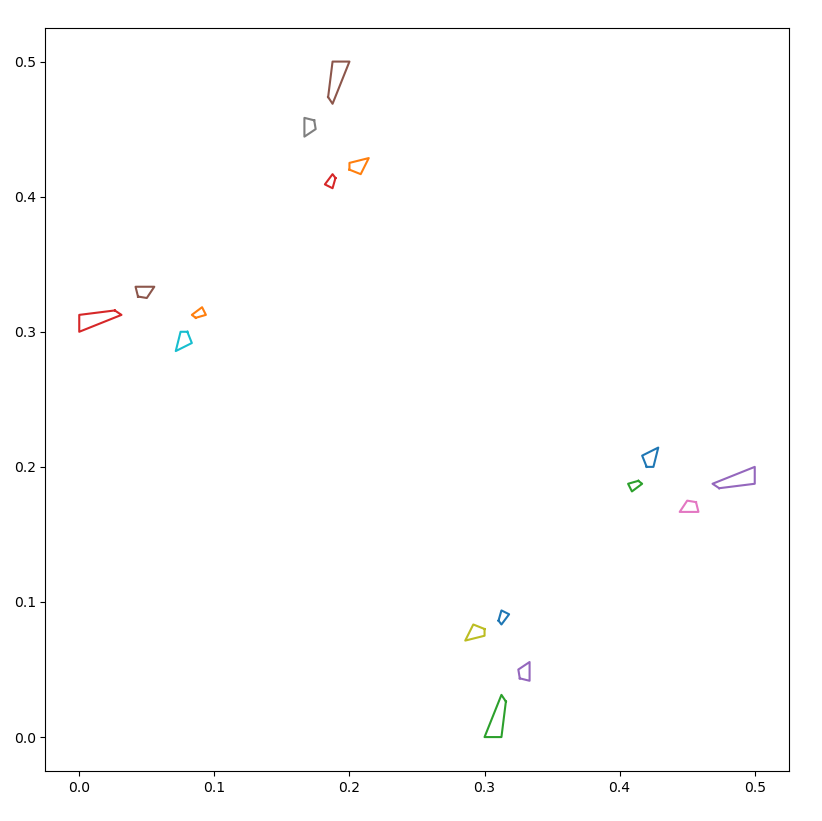}}}%
        \subfloat[$8$ iterations]{{\includegraphics[width=.24\textwidth]{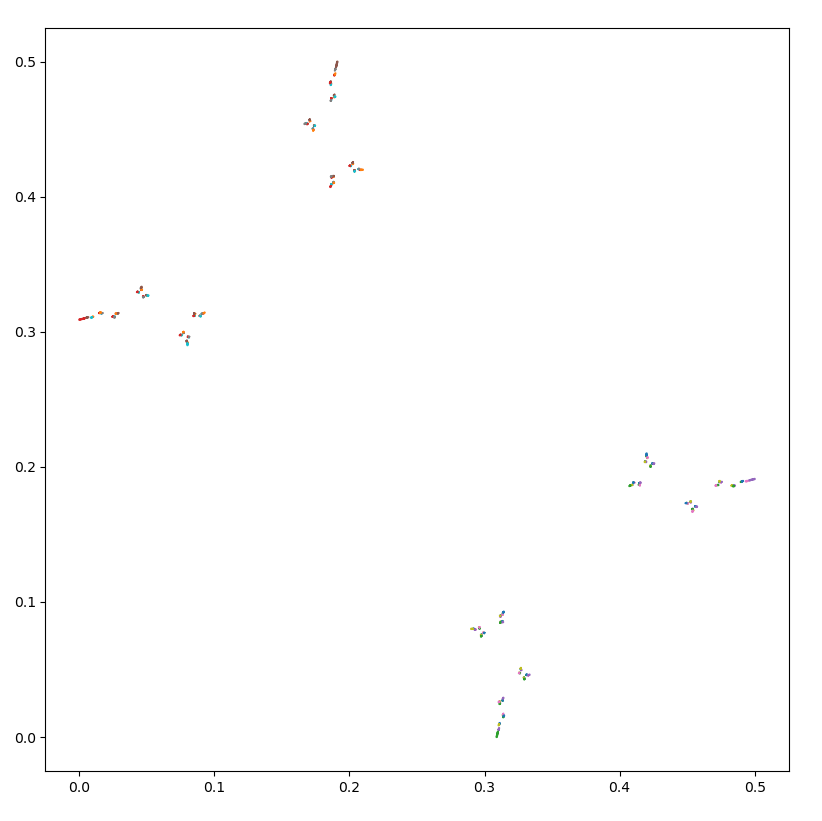}}}
        \caption{Successive approximations of the set $\left\{\left(a(\tau),b(\tau)\right) \;|\; \tau \in \{0,1\}^\mathbb{N}\right\}$ obtained by computing the images of the square $K$ by sequences of iterations of the two homographies $h_0$ and $h_1$.\label{fig:approx-faandb}}
    \end{figure}

\begin{figure}[!ht]
    \centering
    \includegraphics[width=.57\textwidth]{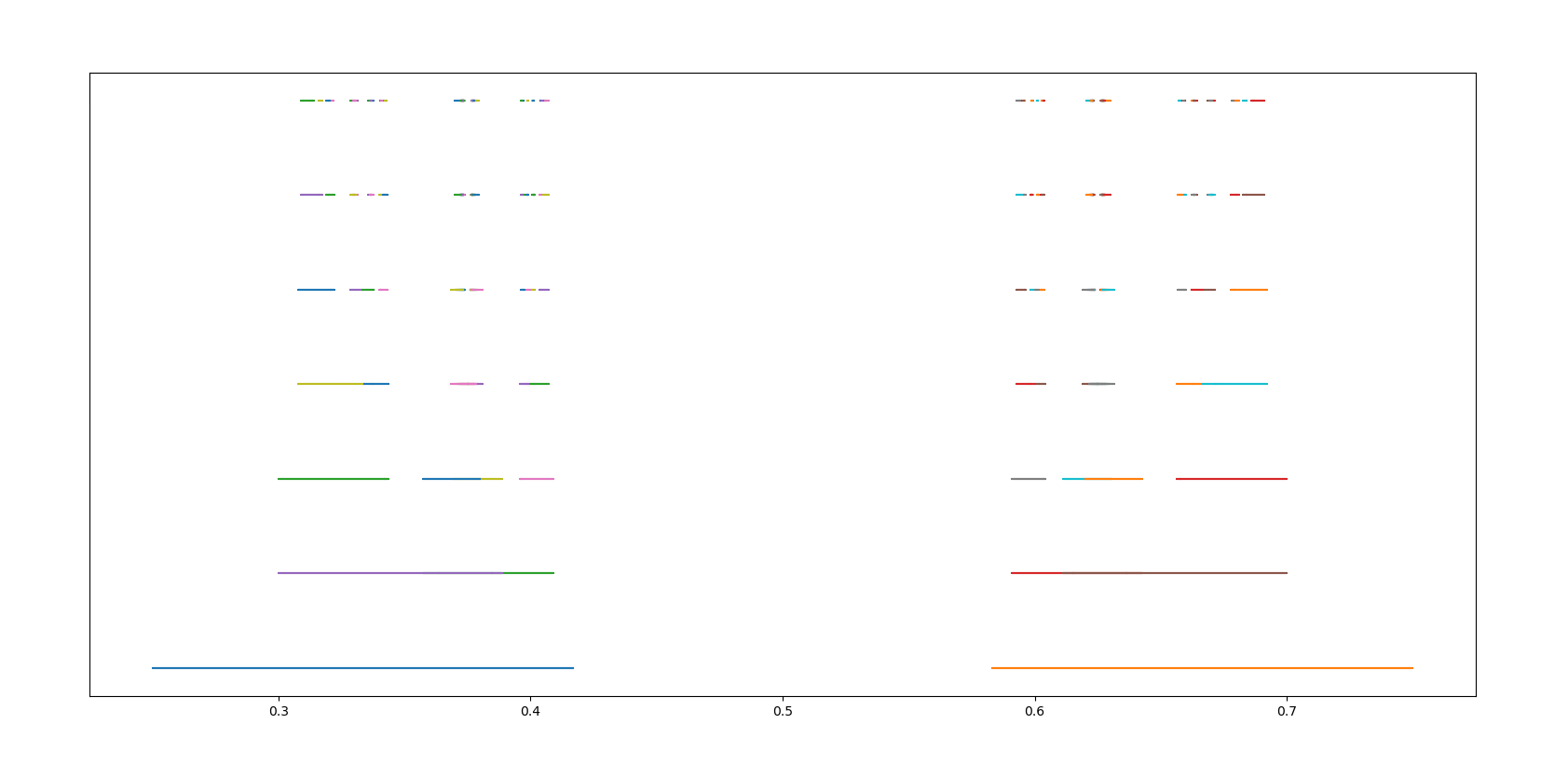}
    \caption{Successive approximations of $F$ (from bottom up), up to rank $7$.\label{fig:approx-f1}}
\end{figure}

Let $F := \left\{f_{\tau} \mid \tau \in \{0,1\}^\mathbb{N}\right\}$ be the set of possible values for the frequency of $1$s in smooth sequences.  
Figure~\ref{fig:approx-f1} illustrates the successive approximations of $F$ derived from the approximations of Figure~\ref{fig:approx-faandb}. Note that $F$ is symmetrical with respect to $1/2$, since swapping the $1$s and the $3$s in a sequence results in moving from a frequency $f$ to a frequency $1-f$. 

When the sequence $\tau$ is ultimately periodic, the value of $f_{\tau}$ has an explicit description. In particular, for $\tau_1=0001^\infty$, we obtain that $f_{\tau_1}=\frac{1}{2} - \frac{1}{13-\sqrt{5}}$, and for $\tau_2=1^\infty$, $f_{\tau_2}=\frac{1}{2} + \frac{2}{3+\sqrt{5}}$. We conjecture that these sequences achieve the maxima of $F$ on the intervals $[0,1/2]$ and $[0,1]$.
    
\begin{problem} Examine in further detail the fractal structure of the set $F$ of possible values for the frequency of $1$s in smooth sequences. Is it true that 
$\sup F\cap [0,1/2] =\frac{1}{2} - \frac{1}{13-\sqrt{5}}$,
and 
$\sup F= \frac{1}{2} + \frac{2}{3+\sqrt{5}}$ ?
\end{problem}
    
\paragraph{Generalisation to other alphabets --} The bi-infinite smooth sequence and recoding formalisms can be extended to any two-integers alphabet $\{\alpha,\beta\}$ with $\A = \alpha^\alpha, \B = \beta^\alpha, \C = \alpha^\beta$ and $\D = \beta^\beta$. 
When both $\alpha$ and $\beta$ are odd integers, the properties of Section~\ref{sec:def} still hold, with new substitutions $\varphi_0$ and $\varphi_1$.

\begin{figure}[!ht]
        \begin{center}
        \subfloat[\centering $\{\alpha,\beta\} = \{1,9\}$\label{fig:approx-1-9}]{{\includegraphics[width=.26\textwidth]{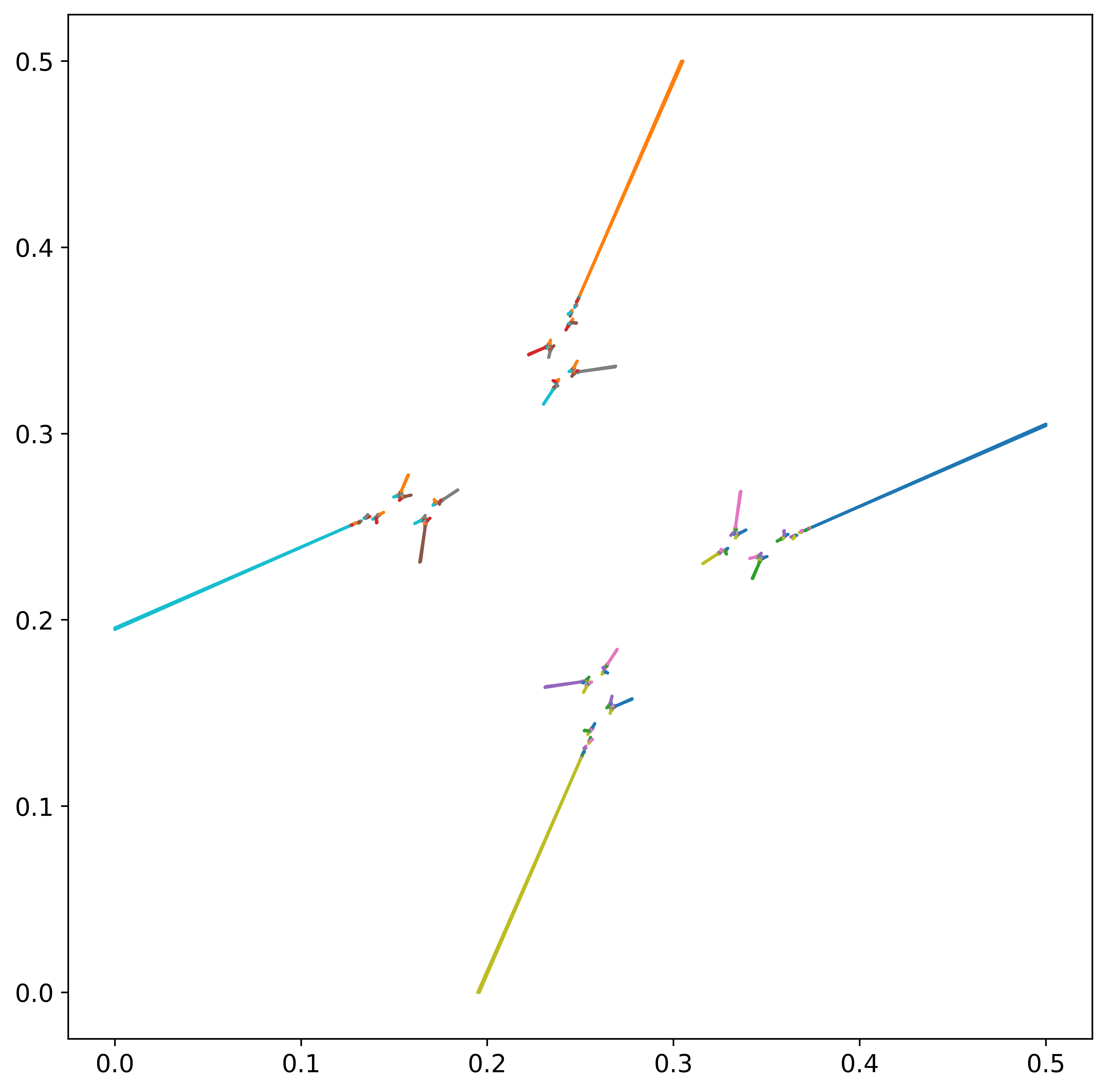}}}%
        \quad
        \subfloat[$\{\alpha,\beta\} = \{1,17\}$\label{fig:approx-1-17}]{{\includegraphics[width=.26\textwidth]{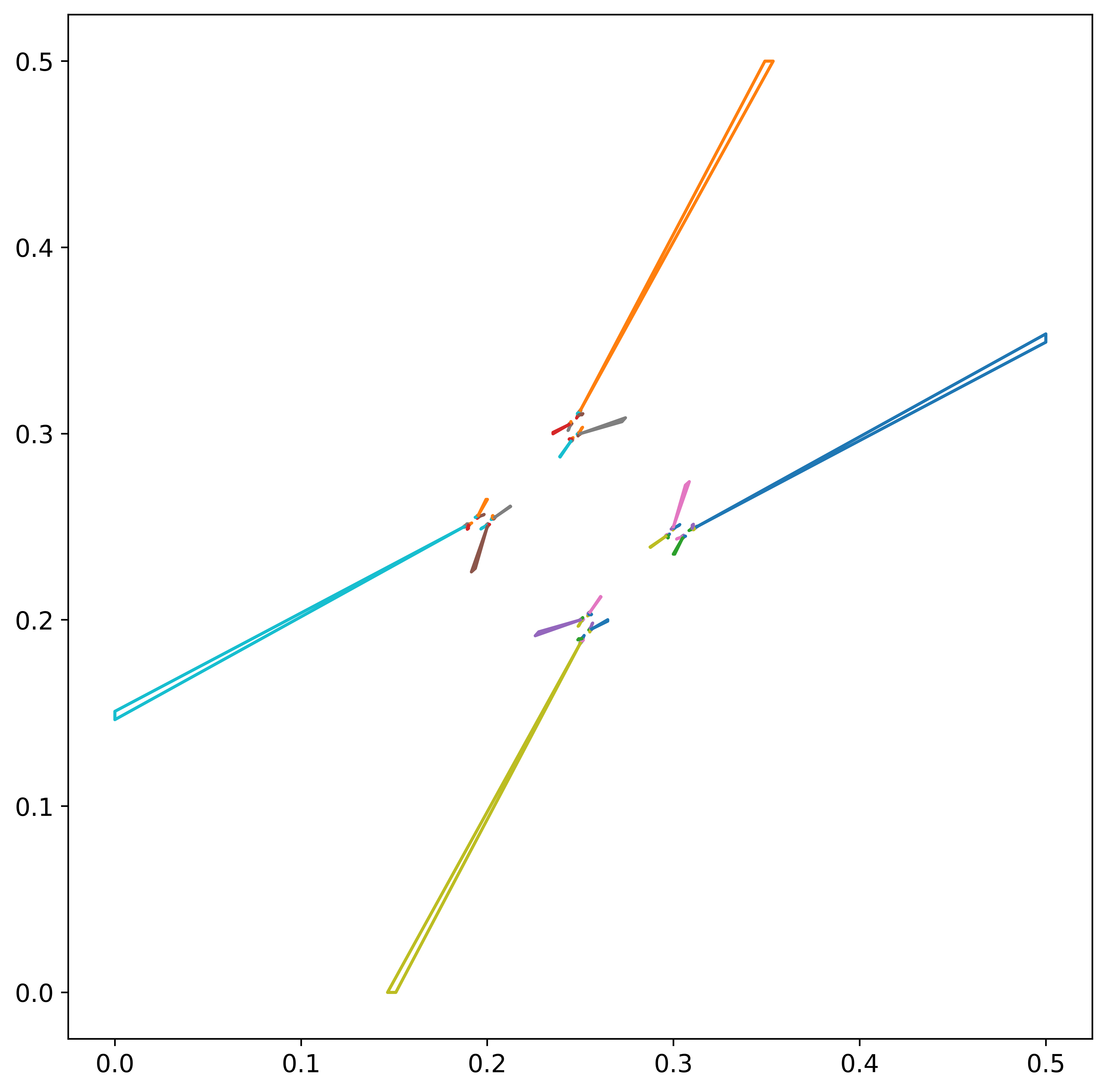}}}
        \\
        \subfloat[\centering $\{\alpha,\beta\} = \{3,7\}$\label{fig:approx-3-7}]{{\includegraphics[width=.26\textwidth]{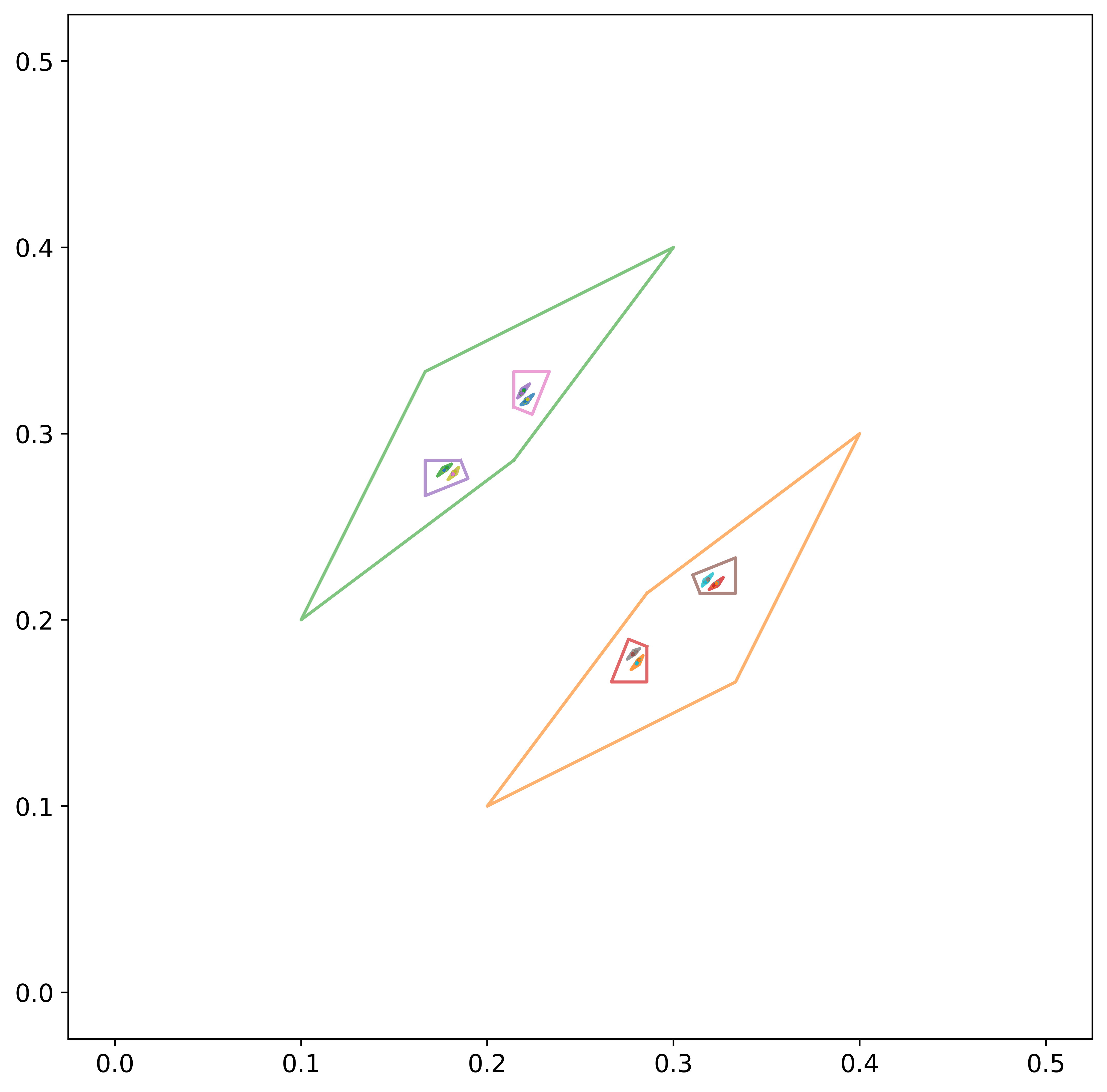}}}%
        \quad
        \subfloat[$\{\alpha,\beta\} = \{9,11\}$\label{fig:approx-9-11}]{{\includegraphics[width=.26\textwidth]{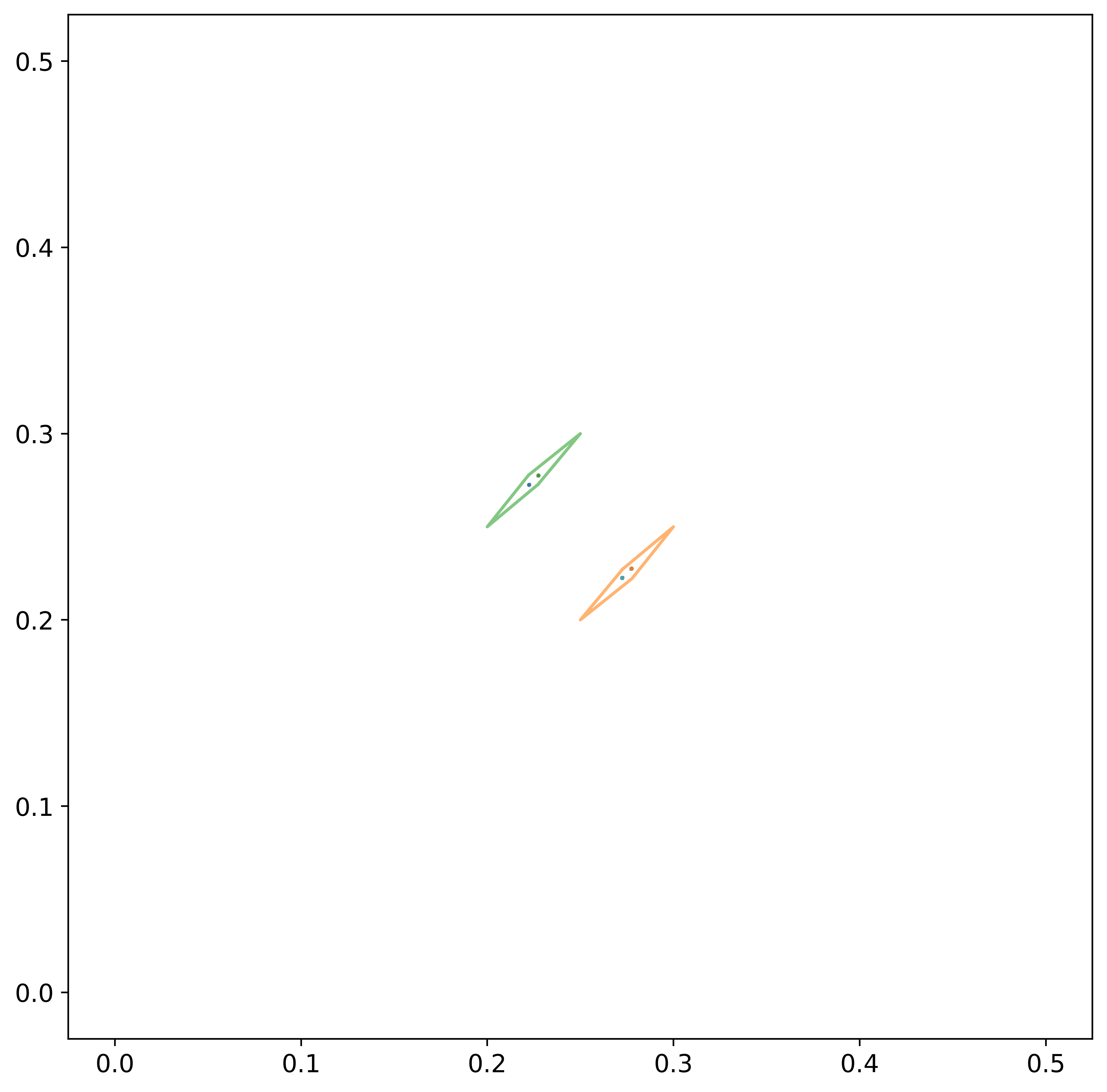}}}
        \caption{Images of the square $K$ by sequences (of length $13$ for (a) and (b), and up to length $6$ for (c) and (d)) of iterations of the two homographies $h_0$ and $h_1$ associated with the given alphabet.\label{fig:approx-other-alphabet-non-contracting}}
        \end{center}
    \end{figure}

However, if we want to use the same arguments as in Section~\ref{sec:MainResults} to prove the unique ergodicity, we need to ensure that the corresponding homographies are contracting on a suitable domain. But this does not seem to be straightforward,  depending on the values of $\alpha$ and $\beta$. Indeed, finite approximations suggest that,  for alphabets of the form $\{1,\beta\}$ with $\beta \geq 5$, the contraction property of the homographies might no longer be valid (see Figures~\ref{fig:approx-1-9} and~\ref{fig:approx-1-17}).

On the other hand, when $\min\{\alpha,\beta\} \ge 3$, the homographies seem to contract event faster than for $\{1,3\}$. See for example Figures~\ref{fig:approx-3-7} and~\ref{fig:approx-9-11} where we had to left the traces of the first iterations since the contraction was happening really fast.

\begin{problem} For which values of $\alpha$ and $\beta$ can we use a contracting property of the homographies $h_0$ and $h_1$ to prove the unicity of $a(\tau)$ and $b(\tau)$? Are the subshifts $(X_{\tau})_{\tau\in\{0,1\}^{\mathbb N}}$ of smooth bi-infinite sequences on the alphabet $\{\alpha, \beta\}$ uniquely ergodic for \emph{any} choice of odd integers $\alpha$ and $\beta$?
\end{problem}


\end{document}